\documentclass[traditabstract,printer]{aa}

\usepackage{graphicx}
\usepackage{txfonts}

\usepackage{bm}
\usepackage{subfigure}
\usepackage{booktabs}

\usepackage[title]{appendix}
\usepackage{enumitem}

\usepackage{color,soul}

\usepackage{natbib,twoopt}
\usepackage{hyperref} 
\bibpunct{(}{)}{;}{a}{}{,}

\usepackage{amstext}

\usepackage{ulem}

\newcommand{\LA}{\left\langle}
\newcommand{\RA}{\right\rangle}

\begin{document}

\pagestyle{plain}
        
        \title{Gravitational Brownian motion as inhomogeneous diffusion: \\
                Black hole populations in globular clusters}
        \titlerunning{Gravitational Brownian motion as inhomogeneous diffusion}

        \author{Zacharias Roupas\inst{\ref{inst1}}}

                \institute{Centre for Theoretical Physics, The British University in Egypt, Sherouk City 11837, Cairo, Egypt \label{inst1}}
                
                \abstract{
                        Recent theoretical and numerical developments supported by observational evidence strongly suggest that many globular clusters host a black hole (BH) population in their centers. This stands in contrast to the prior long-standing belief that a BH subcluster would evaporate after undergoing core collapse and decoupling from the cluster. In this work, we propose that the inhomogeneous Brownian motion generated by fluctuations of the
stellar gravitational field may act as a mechanism adding a stabilizing pressure to a BH population. 
                        We argue that the diffusion equation for Brownian motion in an inhomogeneous medium with spatially varying diffusion coefficient and temperature, which was first discovered by Van Kampen, also applies to self-gravitating systems. 
                        Applying the stationary phase space probability distribution  to a single BH immersed in a Plummer globular cluster, we infer that it may wander as far as $\sim 0.05,\,0.1,\,0.5{\rm pc}$ for a mass of $m_{\rm b} \sim 10^3,
                        \,10^2,\,10{\rm M}_\odot$, respectively.
                        Furthermore, we find that the fluctuations of a fixed stellar mean gravitational field  are sufficient to stabilize a BH population above the Spitzer instability threshold. Nevertheless, we identify an instability whose onset depends on the Spitzer parameter, $S =       (M_{\rm b}/M_\star) (m_{\rm b}/m_\star)^{3/2} ,$ and parameter $B = \rho_{\rm b}(0) (4\pi r_c^3/M_b)(m_\star/m_{\rm b})^{3/2} $, where $\rho_{\rm b}(0)$ is the
                        Brownian population central density. For a Plummer sphere, the instability occurs at $(B,S) = (140,0.25)$. For $B > 140,$ we get very cuspy BH subcluster profiles that are unstable with regard to the support of fluctuations alone. For $S > 0.25,$  there is no evidence of any stationary states for the BH population  based on the inhomogeneous diffusion equation. }
                
                \date{}
                
                \keywords{stellar clusters, gravitational kinetic theory, black holes}       
                
                \maketitle
                
\section{Introduction}

The study of Brownian motion was introduced in astrophysics by \cite{1943RvMP...15....1C}, who used it to establish,  among other things, the concept of dynamical friction  \citep{1943ApJ....97..255C}. Brownian motion has also been used to model or estimate the motion of a massive black hole in the center of a stellar cluster or galaxy  \citep{2002ApJ...572..371C,2002PhRvL..88l1103C, 2003ApJ...592...32C, 2005ApJ...628..673M,2007AJ....133..553M,merritt2013dynamics,2016MNRAS.461.1023B,2018MNRAS.473.1719L,2020IAUS..351...93D}. 
While significant developments have been achieved regarding the statistical mechanics and a general kinetic theory of orbit-averaged motion in action-angle variables \citep{1984MNRAS.209..729T,1988MNRAS.230..597B,Chavanis_2012,2013A&A...556A..93C,Vasiliev_2017,2017ApJ...842...90R,2018MNRAS.481.4566F,2020MNRAS.493.2632T} or other canonical variables \citep{2020JPhA...53d5002R}, a diffusion equation in configuration space for inhomogeneous Brownian motion -- including varying velocity dispersion and diffusion coefficient-- has not been applied as of yet. The paradigm of inhomogeneous Brownian motion is directly relevant to a self-gravitating system consisting of a population of heavier, fewer bodies immersed in a bigger, highly populated cluster of lighter bodies, whose profile is typically inhomogeneous regarding the density, velocity dispersion, and diffusion coefficient.

Such a population of a significant number of black holes (BHs) is believed to exist in the center of many globular clusters. Its existence is supported by numerical and theoretical developments \citep{2004ApJ...608L..25M,2008MNRAS.386...65M,2013ApJ...763L..15M,2013MNRAS.432.2779B,2015ApJ...800....9M,2016MNRAS.458.1450W,2016MNRAS.455...35A,2016MNRAS.463.2109R,2017ApJ...834...68C,2018ApJ...855L..15K,2018MNRAS.478.1844A,2018MNRAS.479.4652A,2018ApJ...864...13W,2019arXiv191109125W,2019ApJ...871...38K}
as well as observational evidence \citep{2007Natur.445..183M,2008ApJ...689.1215B,2012Natur.490...71S,2010ApJ...712L...1I,2012ApJ...760..135R,2013ApJ...777...69C,2015MNRAS.453.3918M,2015ApJ...805...65T,2015ApJ...810L..20M,2017MNRAS.467.2199B,2018MNRAS.475L..15G,2018ApJ...855...55S,2019ApJ...884L...9A}. These recent advances stand in contrast to a long-held belief that globular clusters cannot retain their BHs \citep{1969ApJ...158L.139S,1993Natur.364..421K,1993Natur.364..423S}. Spitzer instability \citep{1969ApJ...158L.139S} seems unavoidable for massive stars which are subject to mass segregation, decoupling from the cluster and undergoing gravothermal collapse. The resulting black hole subcluster, according to a traditional view, would itself undergo core collapse and become so dense that would evaporate due to two-body and three-body encounters, apart from one or two BHs that may be retained in the centre \citep{1993Natur.364..421K,1993Natur.364..423S}. However, examining this picture more carefully, we may realize that since Spitzer instability depends on total mass and the BH-population has significantly less total mass than the progenitor subcluster of massive stars, the former might not undergo gravothermal collapse and might, instead, become stabilized. It has also been  suggested that the BH-population does not stay decoupled from the cluster for as long a time as initially thought \citep{2013MNRAS.432.2779B,2013ApJ...763L..15M}.

Here, we propose an additional theoretical component to this picture. We investigate whether a BH-population that is well within the Spitzer instability regime may be supported by Brownian motion induced by the fluctuations of the cluster's gravitational field. 
To this end, we describe the Brownian motion of the BHs due to fluctuations of the field by an inhomogenous diffusion equation that takes into account the density, velocity dispersion, and dynamical friction coefficient spatial variations.  This model for the inhomogeneous diffusion of Brownian particles was discovered by  \cite{1988JPCS...49..673V} in a general setting and we argue that it also applies to self-gravitating systems.

In the next section, we review the Van Kampen inhomogeneous diffusion equation for Brownian particles and its stationary solution. In Section \ref{sec:IMBH}, we study a single massive BH and in Section \ref{sec:BH_sub}, we consider a BH population immersed in a fixed Plummer profile. In the final section, we discuss our conclusions. In Appendix \ref{app:diffusion}, we re-derive the Van Kampen diffusion equation.

\section{Inhomogeneous diffusion of black holes in stellar clusters}

A standard approach in gravitational kinetic theory is to apply the orbit-averaged Fokker-Planck equation (e.g. 
\citealt{merritt2013dynamics,B&T_2008gady.book}). 
This approach comes with a drawback that \cite{merritt2013dynamics}\footnote{In page 253, section 5.5.} refers to as a ``kludge.'' He remarks that while the diffusion coefficients are averaged over the volume filled by an orbit, no account is taken of the fact that the perturbations acting on a test star come from stars in regions far from the test star, which may have very different properties than those near to the test star. For this reason he concludes there is no sense in which the orbit-averaged Fokker-Planck equation might   correctly describe an inhomogeneous system. We wish to describe the motion of heavier -- that is, Brownian -- bodies inside an inhomogeneous (in phase-space) medium of lighter bodies and, in addition, we wish to focus on the distribution of position. Therefore, we shall not use the standard orbit-averaged approach with $E$, $L$ degrees of freedom, but the older approach from \cite{1943ApJ....97..255C,1943RvMP...15....1C} in the phase-space $x$, $v$ (see also section 5.4 of \citealt{merritt2013dynamics}). We nevertheless generalize \cite{1943RvMP...15....1C} approach here in order to account for inhomogeneity in space and temperature.

This may be achieved by considering the Kramers equation of the Fokker-Planck type \citep{1940Phy.....7..284K} in a local sense (Equation \ref{eq:Kramers}) and derive a diffusion equation in a way that is similar to the derivation of the Smolukowski equation from the Kramers equation, as done by \cite{1943RvMP...15....1C}. This problem was already resolved  by \cite{1988JPCS...49..673V} and we provide a review in Appendix \ref{app:diffusion}.

Therefore, we suggest that the Van Kampen diffusion equation for Brownian motion in inhomogeneous media \citep{1988JPCS...49..673V} applies to self-gravitating systems, such as stellar clusters, which is the focus of this work, as well as dark matter haloes. These are typically spatially inhomogeneous and non-isothermal and they acquire a velocity dispersion that is also varying. If we denote $\mu(x) \equiv 1/m\eta(x),$ with $\eta(x)$ as the diffusion coefficient and $m$ the mass of a Brownian particle, $\Phi(x)$ and $T(x)$ the potential and local temperature of the system in which the Brownian particle is embedded and $p(x)$ the probability density of the Brownian particle in one dimenion $x$, the Van Kampen diffusion equation reads:
\begin{equation}\label{eq:Brownian_diff}
\frac{\partial p (x,t)}{\partial t} = \frac{\partial }{\partial x} \left\lbrace \mu (x)  \left( m\Phi(x)^\prime p(x,t)  + \frac{\partial}{\partial x}(T(x) p (x,t)) \right)\right\rbrace.
\end{equation}
As we noted above, we reproduce this equation in Appendix \ref{app:diffusion}. To be more specific, it is derived by expanding the Fokker-Planck equation (\ref{eq:Kramers} of Kramers with respect to $\eta^{-1}$, defined in the Langevin equation (\ref{eq:Langevin}).  This quantity is given in units of time and corresponds to the timescale of the diffusion, as was suggested by \cite{1943ApJ....97..255C}. Therefore, the Van-Kampen equation is strictly valid for times $t\gg \eta^{-1}$. 

The stationary distribution 
\begin{equation}
\frac{\partial p}{\partial t} = 0
\end{equation}
of Equation \ref{eq:Brownian_diff} is  
\begin{equation}
p(x) = \frac{p(0)}{\beta (0)}\beta(x) e^{-\int_{0}^{x} d\tilde{x}\,\beta (\tilde{x}) m\Phi(\tilde{x})^\prime},
\end{equation}
where it is assumed that $p(0)'=T(0)'=\Phi(0)'=0$. We denoted $\beta = 1/T$.
In three dimensions, the diffusion Equation \ref{eq:Brownian_diff} becomes
\begin{equation}\label{eq:Brownian_diff_3D}
\frac{\partial p (\bf{r},t)}{\partial t} = \nabla \left\lbrace \mu ({\bf{r}}) \left( m (\nabla \Phi ({\bf{r}}))  p ({\bf{r}},t)   + \nabla( T({\bf{r}}) p ({\bf{r}},t) ) \right)\right\rbrace.
\end{equation}
In the spherically symmetric case, the stationary radial probability density can therefore be written as
\begin{equation}\label{eq:p_stationary}
p(r) = \frac{ \beta(r) e^{-\int_{0}^{r} d\tilde{r}\, \beta (\tilde{r}) m \Phi(\tilde{r})^\prime } }{\int_0^\infty dr\, 4\pi r^2 \beta(r) e^{-\int_{0}^{r} d\tilde{r}\, \beta (\tilde{r}) m\Phi(\tilde{r})^\prime } }.
\end{equation}
The result is that at equilibrium, the probability density does not depend on $\mu(r)$, but only on $\beta(r)$ and $\Phi(r)$.
We note that the stationary phase space distribution function may be written at first order in $\eta(x)^{-1}$ by use of Equations (\ref{eq:f0}), (\ref{eq:f1}), and (\ref{eq:n_exp}):\ 
\begin{align}\label{eq:f_stat}
f(x,v) &= p(x) \sqrt{\frac{m}{2\pi T(x)}} e^{-\frac{mv^2}{2T(x)}} \nonumber \\
&\left\lbrace 
1 - \eta(x)^{-1}\left\lbrace v\left(\frac{1}{2}\frac{T(x)^\prime}{T(x)} p(x) + p(x)^\prime \right) + v^3\frac{m T(x)^\prime}{T(x)^2}p(x)
\right\rbrace
\right\rbrace.
\end{align}
The temperature, $T(x),$ refers to the host cluster. There  a small correction term arises to the Maxwellian distribution. This requires further investigation and may be the subject of  interesting future developments, but it is not discussed in this work.

\subsection{Single black hole}\label{sec:IMBH}

\begin{figure}[!tb]
        \begin{center}
                \subfigure[]{
                        \label{fig:p_IMBH}
                        \includegraphics[scale = 0.5]{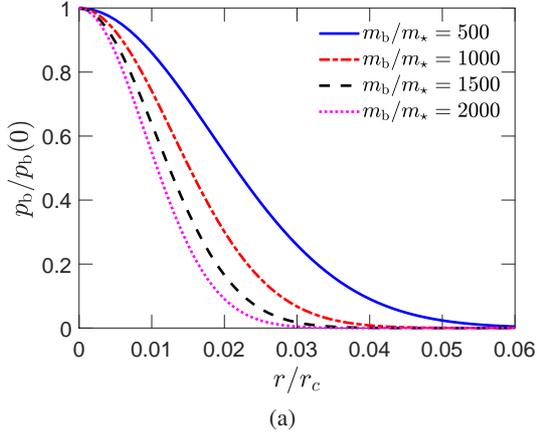}  }
                \subfigure[]{
                        \label{fig:p_IMBH}
                        \includegraphics[scale = 0.5]{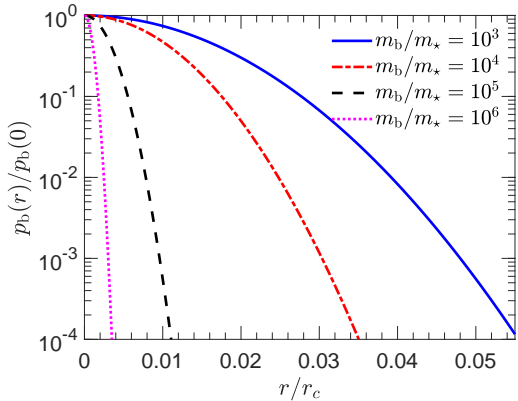}  }
                \caption{Stationary probability distribution $p_{\rm b}$ of inhomogeneous diffusion (\ref{eq:p_stationary}) with respect to distance, $r,$ for a single Brownian body, $m_{\rm b}$, inside a Plummer external gravitational potential. The Brownian body may be considered to be a massive  BH and the gravitational potential may be generated by a globular cluster of individual average stellar mass, $m_\star$. We denote $r_c$ the softening radius of the potential.}
                \label{fig:IMBH}
        \end{center}
\end{figure}

 Here, we consider a single BH of mass, $m_{\rm b}$, as a Brownian particle immersed inside a stellar cluster of total mass, $M_\star,$ and average individual stellar mass, $m_\star$.

We model the distribution of the stellar cluster with a Plummer sphere,
\begin{align}\label{eq:Plummer}
\rho_\star (x) &= \frac{M_\star}{\frac{4}{3}\pi r_c^3}\frac{1}{(1+x^2)^{5/2}}
,\;
\mathcal{M}_\star (x) = M_\star \frac{x^3}{(1+x^2)^{3/2}}
,\\
\sigma_\star (x)^2 &= \frac{G M_\star}{6r_c}\frac{1}{(1+x^2)^{1/2}},
\end{align}
where $\rho_\star(x)$, $\sigma_\star (x)$ denote the mass density and velocity dispersion of the host cluster, respectively, while $\mathcal{M}_\star(x)$ denotes the total stellar mass contained within $x$.
The half-mass radius is related to the softening radius $r_c$ by
\begin{equation}
r_{\rm hm} = (2^{2/3} - 1)^{-1/2} r_c \simeq 1.3 r_c.
\end{equation}
We identify the temperature as
\begin{equation}
T_\star (r) = m_\star \sigma_\star(r)^2.
\end{equation}

The probability density of the BH position is given in a straightforward way from Equation (\ref{eq:p_stationary}) which, in the case of Plummer external potential, may be written as
\begin{equation}\label{eq:p_stationary_single}
p_{\rm b} = p_{\rm b}(0) (1+x^2)^{-\frac{1}{2}-3\frac{m_{\rm b}}{m_\star}},
\end{equation}
where 
\begin{equation}
p_{\rm b}(0) = \frac{1}{4\pi r_c^3} \frac{4\Gamma\left(3\frac{m_{\rm b}}{m_\star} + \frac{1}{2} \right)}{\sqrt{\pi}\Gamma\left(3\frac{m_{\rm b}}{m_\star} - 1 \right)},
\end{equation}
and $\Gamma$ is the gamma-function. In Figure \ref{fig:IMBH}, we plot the probability density with respect to $x$ for several different values of the individual mass ratio $m_{\rm b}/m_\star$. For a dense stellar cluster, such as a globular cluster or a nuclear star cluster, it is $r_c\sim 1-5{\rm pc}$ and $m_\star \sim0.5{\rm M}_\odot$. It is evident that an intermediate-mass BH with $m_{\rm b} \sim 10^3{\rm M}_\odot$ inside a globular cluster wanders as far as $\sim 0.05 {\rm pc}$ from the center, and in a nuclear star cluster, as far as $\sim 0.2 {\rm pc}$, although a Plummer sphere might not be a fairly good approximation for the density of the latter. A steeper external profile would induce smaller diffusion and, therefore, stricter boundaries for the BH fluctuation.            

\begin{figure}[!tb]
        \begin{center}
                \subfigure[Brownian motion of a BH.]{
                        \label{fig:simulation_rt_single_run_N=1e3}
                        \includegraphics[scale = 0.5]{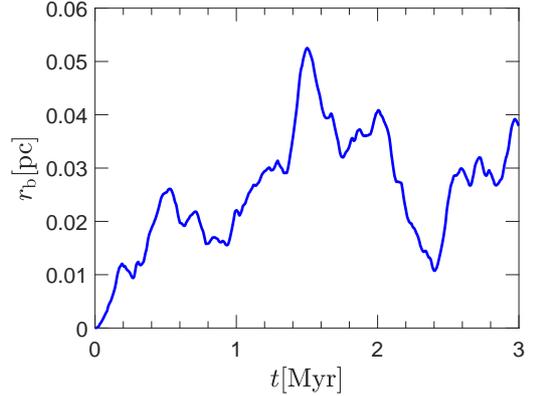}  }
                \subfigure[Probability density of the position of a BH.]{
                        \label{fig:simulation_dPdV_N=5e2_relax_IC}
                        \includegraphics[scale = 0.5]{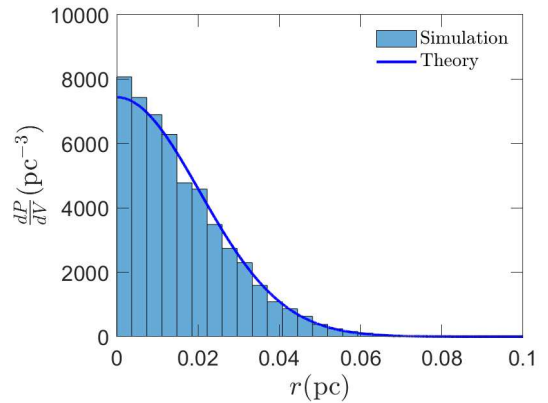}  }
                \caption{$N$-body simulations with AMUSE framework of a BH with mass $m_{\rm b} = 50 {\rm M}_\odot$ immersed with zero velocity in the center, $r_{\rm b, ini} = 1{\rm AU,}$ of a stellar cluster with $N_\star=1000$, $m_\star=0.5{\rm M}_\odot$, and initial conditions sampled from a Plummer sphere with softening radius, $r_c = 0.5{\rm pc}$. (a) Radial position of the BH with respect to time in a single run. (b) Probability density of the position of the BH for 1000 runs. See text for details.}
                \label{fig:simulation_dPdV}
        \end{center} 
\end{figure}

In order to verify, in a straightforward manner, the validity of the inhomogeneous diffusion Equation (\ref{eq:Brownian_diff}), we performed direct numerical $N$-body simulations using the AMUSE 13.2 framework \citep{2009NewA...14..369P,2013CoPhC.184..456P,2013A&A...557A..84P,10.1088/978-0-7503-1320-9}. We considered a stellar cluster with $N_\star=10^3$, $m_\star=0.5{\rm M}_\odot$, $r_c = 0.5{\rm pc}$ and a BH with $m_{\rm b}=50{\rm M}_\odot$ placed in the center of the cluster at distance $r_{\rm b, ini} = 1{\rm AU}$ and zero velocity. We use this initial condition for the BH to demonstrate that even if it is left sitting still in the center, it will eventually be kicked out due to gravitational fluctuations. In any case, we find that the distribution function seems to not be affected by the BH initial conditions, at least when it is initially bound inside the core $\sim 0.6 r_c$. We performed 1000 runs, initially sampling  the stellar cluster from an isotropic Plummer phase-space distribution function $f_{\rm Pl}(x,v)\sim (-E(x,v))^{7/2}$. In our previous calculation (Equation \ref{eq:p_stationary_single}), we assumed the Plummer potential to be fixed. 
Therefore, our sampling of the BH position should occur at a time is shorter than the relaxation time of the cluster $t_{\rm rh} =  0.138 (N_\star/\ln\Lambda) (G\rho_\star)^{-1/2}$ (e.g., \citealt{2003gmbp.book.....H,10.1088/978-0-7503-1320-9})
so that the deviation from Plummer distribution is small. The relaxation time for a Plummer sphere is (Section 8, Table 1 of \citealt{2003gmbp.book.....H}):
\begin{equation}\label{eq:t_rh_Plummer}
        t_{\rm rh} =  \frac{0.206}{\ln\Lambda} \frac{N_\star r_c^{3/2}}{\sqrt{GM_\star}} = 10{\rm Myr},
\end{equation} 
where we used $\Lambda = 0.11 N_\star$ \citep{10.1088/978-0-7503-1320-9}. The BH diffusion timescale $t_{\rm b, rel}$ is  given from the two-body relaxation timescale (in a closely related context see \citealt{2020MNRAS.492..877E,2016MNRAS.461.1745E}):
\begin{equation}\label{eq:t_b_rel}
        t_{\rm b,rel} =  \frac{1}{8\pi \ln\Lambda} \frac{\sigma_\star^3}{G^2 \rho_\star m_{\rm b} } .
\end{equation} 
Dividing the two timescales, we get at the center $r=0:$
\begin{equation}
        \frac{t_{\rm b,rel} }{t_{\rm rh}} = 0.05\frac{m_\star}{m_{\rm b}}.
\end{equation}
This verifies that the diffusion Equation (\ref{eq:Brownian_diff}) is valid at times that are comparable to the relaxation timescale of a cluster for $m_\star < m_{\rm b}$.
Now, for the parameters in our simulation, the last equation gives $t_{\rm b,rel} = 5\cdot 10^{-4}   t_{\rm rh} = 5\cdot 10^{-3}{\rm Myr}$. We acquire the BH position at each simulation run within the time window, $t = 0.3\pm 0.05 {\rm Myr}$, which is sufficiently bigger than $t_{\rm b, rel}$, so that the BH has enough time to diffuse, and sufficiently smaller than $t_{\rm rh}$, so that the Plummer potential does not get drastically altered.

 We calculate the probability density by assuming the ergodic hypothesis holds (at each run, we record several positions in time) and performing an ensemble average. The fit of the theory of inhomogeneous diffusion to the simulations is very good, as depicted in Figure \ref{fig:simulation_dPdV}. In Figure \ref{fig:simulation_rt_single_run_N=1e3}, we demonstrate in a single simulation run that the BH rapidly (at $\sim t_{\rm b,rel}= 0.005 {\rm Myr}$) diffuses in the medium, while in Figure   \ref{fig:simulation_dPdV_N=5e2_relax_IC}, we present the BH probability density of 1000 simulation runs. The RMS position $r_{\rm RMS} \equiv < r_{\rm b}^2>^{1/2}$ of the simulations $r_{\rm RMS}^{\rm sim} = 0.037 {\rm pc}$ is well-matched by the theoretical prediction $r_{\rm RMS}^{\rm theory} = 0.036 {\rm pc}$ in the chosen time window.  
        
\subsection{Black hole subcluster}\label{sec:BH_sub}

\begin{figure}[!tb]
        \begin{center}
                \subfigure[Series of equilibria for various mass ratios.]{
                        \label{fig:M_ratio_p0}
                        \includegraphics[scale = 0.5]{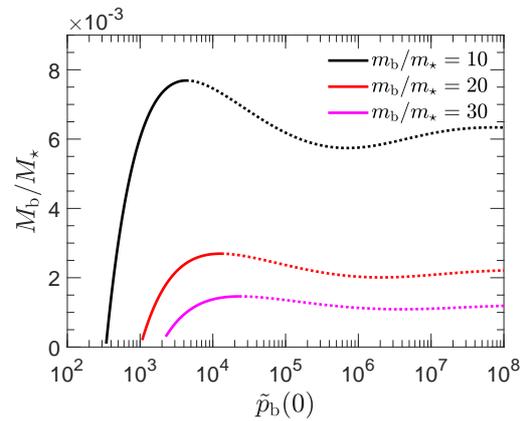}  }
                \subfigure[Global series that applies to any subcluster.]{
                        \label{fig:Spitzer}
                        \includegraphics[scale = 0.5]{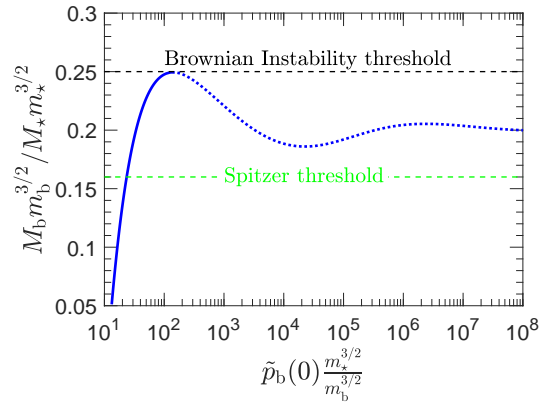}  }
                \caption{(a) Series of equilibria of the Brownian population with total mass $M_{\rm b}$ embedded in a Plummer host cluster with total mass, $M_\star,$ for three different individual mass ratios. The $x$-axis is the dimensionless density at the center. The presence of a maximum at each curve designates an instability. No equilibrium exists above the maximum at each case, while the dotted curves correspond to equilibria that cannot be supported by inhomogeneous Brownian pressure alone; (b) Three curves of (a) merge to a single one when scaled properly with the individual mass ratios. The $y$-axis variable is the Spitzer parameter $S$. In Brownian inhomogeneous diffusion, equilibria exist above the Spitzer instability threshold $S_{\rm Spitzer} = 0.16$. The instability sets in at $S = 0.25$, $\tilde{p}_{\rm b}(0)(m_\star/m_{\rm b})^{3/2} = 140$ in the case of a Plummer host cluster.}
                \label{fig:Spitzer_M}
        \end{center} 
\end{figure}

Here, we consider a two-component model, namely a population of $N_{\rm b}$ Brownian particles with average mass, $m_{\rm b}$ embedded in a cluster of $N_\star$ bodies with average mass, $m_\star$. Our description wishes to describe a BH population immersed in a globular cluster, but applies generally to any self-gravitating system which hosts a subsystem of significantly less bodies. The total mass of the host is $M_\star = N_\star m_\star$ and the total mass of the Brownian bodies is $M_{\rm b} = N_{\rm b} m_{\rm b}$. 

The potential $\Phi$ includes the potential of the host cluster, but also can account for the self-gravity of the Brownian population if not negligible. Given the distribution of the host cluster and neglecting the feedback of the Brownian population onto the distribution of the host, the diffusion equation (\ref{eq:Brownian_diff}) together with the Poisson equation form a system of equations that determines the equilibrium distribution of the Brownian particles. This system may be formulated as follows. In the spherically symmetric case, the gravitational field at any point $r$ may be decomposed according to Poisson equation as
\begin{equation}
        \frac{d\Phi}{dr} = G\frac{\mathcal{M}_\star(r)+\mathcal{M}_{\rm b}(r)}{r^2},
\end{equation}
where $\mathcal{M}_\star$, $\mathcal{M}_{\rm b}$ denote the mass contained in $r$ of the host cluster and the subcluster respectively.
The equilibrium corresponds to the stationary solution of the inhomogeneous diffusion equation (\ref{eq:Brownian_diff}). We therefore get the system for the Brownian population:
\begin{align}
\label{eq:ode_system_d1}
        \frac{dp_{\rm b}(r)}{dr} &= -p_{\rm b}(r)\left( Gm_{\rm b}\frac{\mathcal{M}_\star(r)+\mathcal{M}_{\rm b}(r)}{r^2 T_\star(r)} + \frac{T_\star(r)^\prime}{T_\star(r)} \right)
        ,\\
\label{eq:ode_system_d2}
                \frac{d\mathcal{M}_{\rm b}(r)}{dr} &=
                4\pi r^2 M_{\rm b} p_{\rm b}(r). 
\end{align}
This is a system of the unknown distributions $\{p_b,\mathcal{M}_{\rm b} \}$ given the host cluster distributions $\mathcal{M}_\star(r)$ and $T_\star(r)$ and subject to the constraint $\mathcal{M}_b(\infty) = M_{\rm b}$. The constraint suggests that equilibria may exist only for certain parameter values. This formulation, where the host cluster distribution is fixed, does not take into account the feedback of the Brownian population onto the host. This may be significant in the proximate regions close to the population if it is sufficiently compact, but we do not be consider this point in this work.

\begin{figure*}[!tb]
        \begin{center}
                \subfigure[$N_{\rm b} = 20$, $N_\star=2\cdot 10^5$]{
                        \label{fig:rho_r_GC_mb=10_N=2e5_Sb-point}
                        \includegraphics[scale = 0.35]{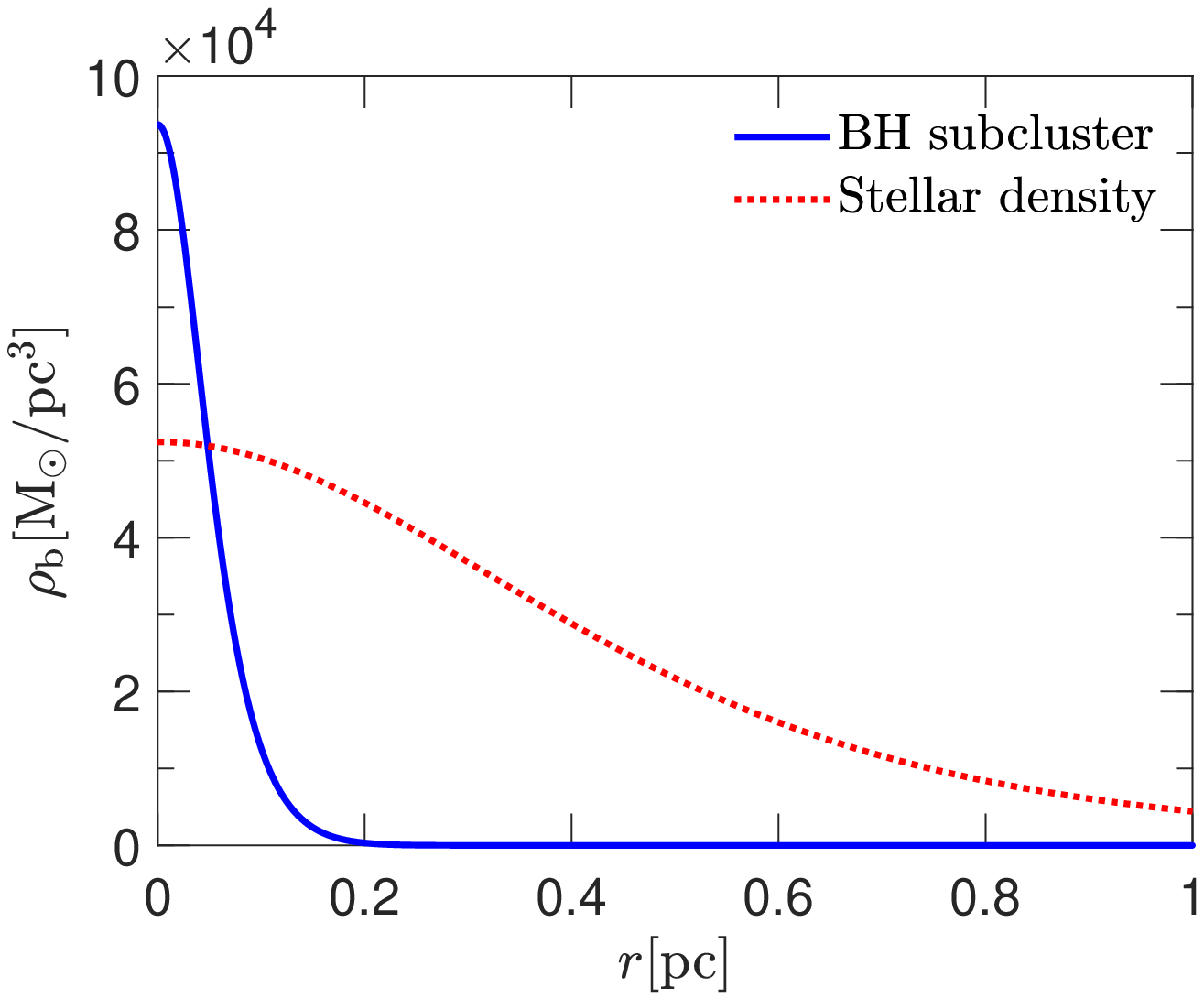}  }
                \subfigure[$N_{\rm b} = 100$, $N_\star = 10^6$]{
                        \label{fig:rho_r_GC_mb=10_N=1e6_Sb-point}
                        \includegraphics[scale = 0.35]{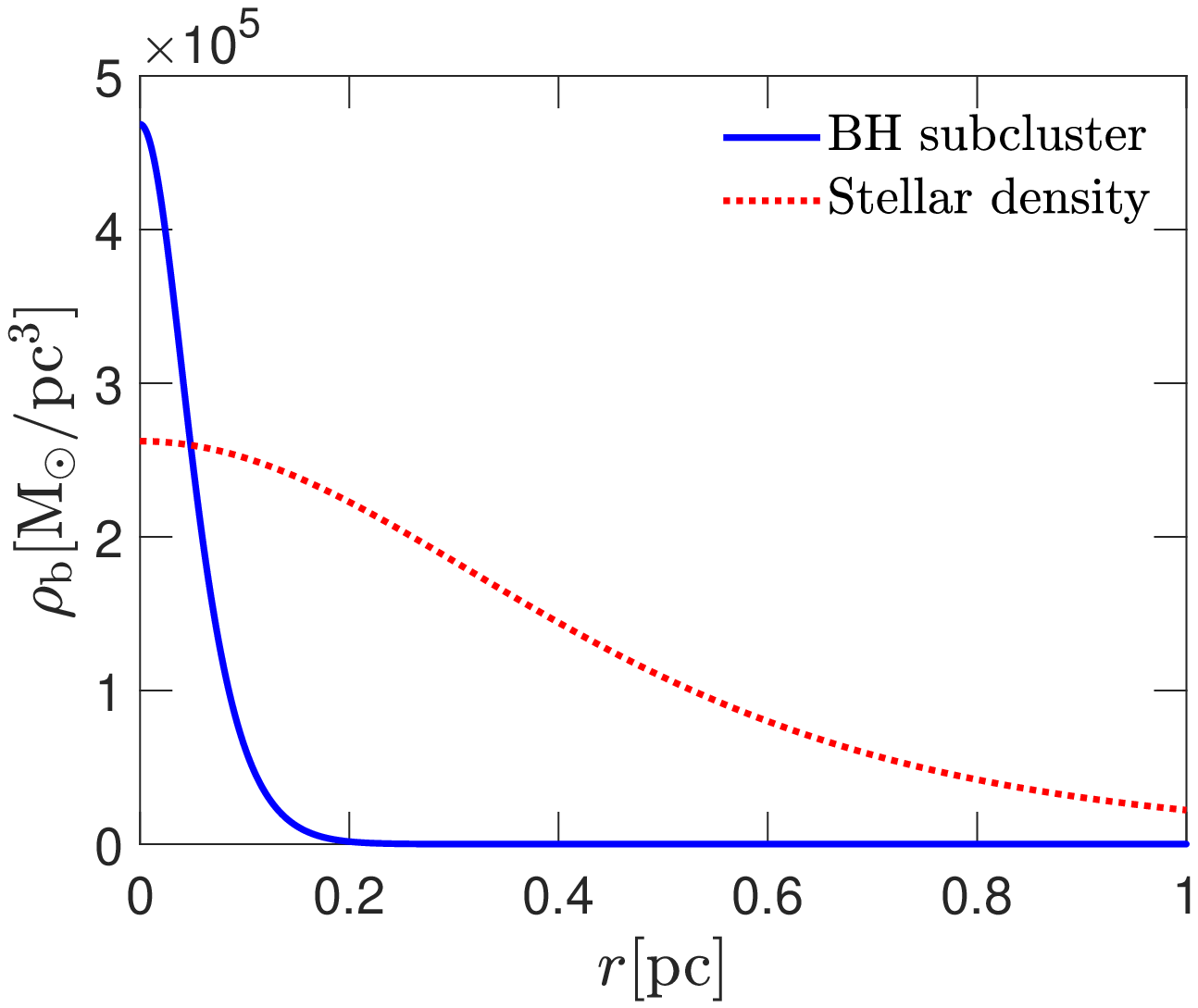}  }
                \subfigure[$N_{\rm b} = 500$, $N_\star=5\cdot 10^6$]{
                        \label{fig:rho_r_GC_mb=10_N=5e6_Sb-point}
                        \includegraphics[scale = 0.35]{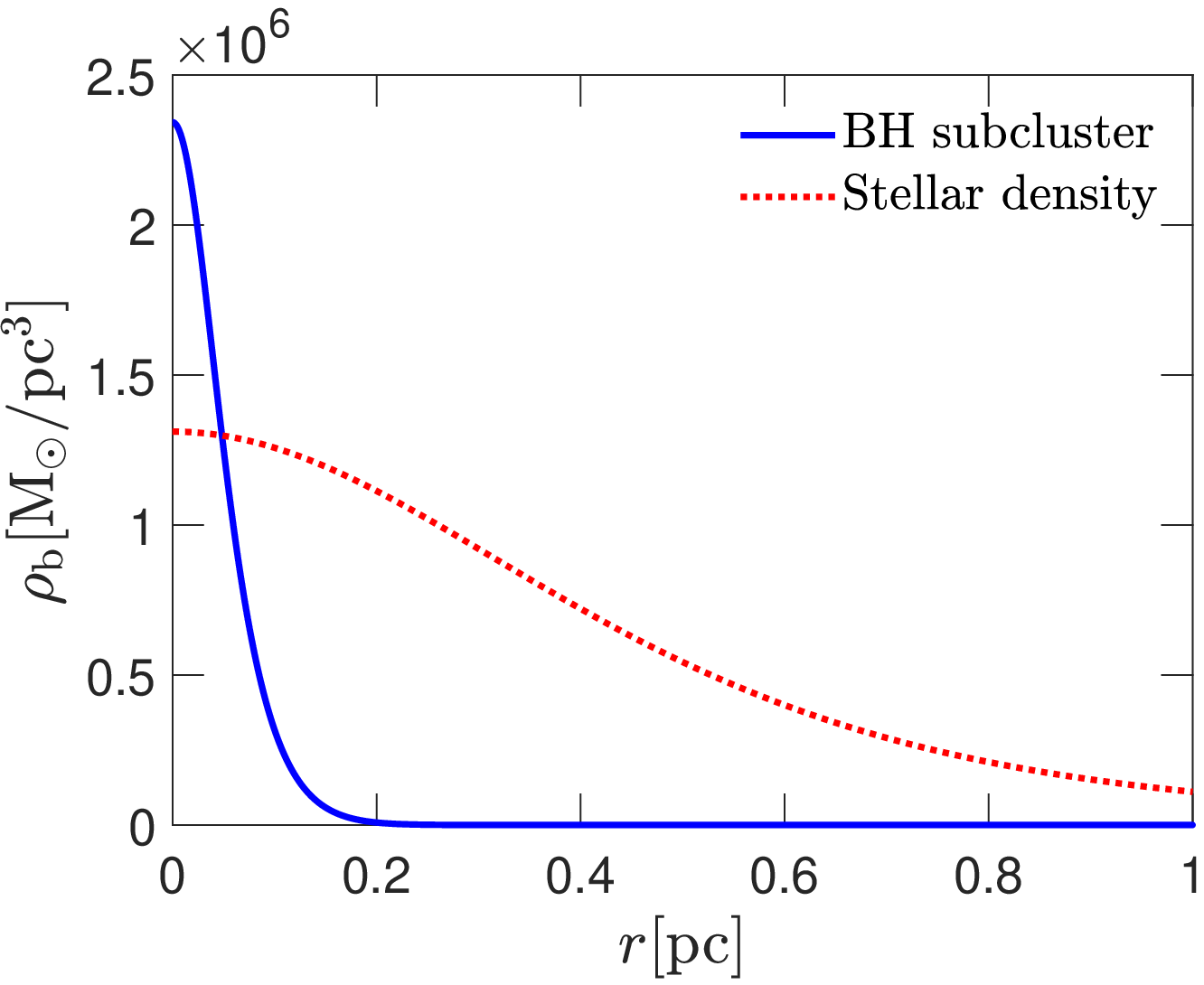}  }
                \caption{Mass density $\rho_{\rm b}(r) = M_{\rm b} p_{\rm b}(r)$ of a marginally stable BH population given by the solution of the system (\ref{eq:ode_system_Pl_1})-(\ref{eq:ode_system_Pl_2}). The average individual BH mass is assumed to be $m_{\rm b} = 10{\rm M}_\odot$ and the Spitzer parameter equals the marginal value of the onset of the instability $S=0.25$ . The BH population is embedded in a globular cluster with average individual mass density $m_\star = 0.5 {\rm M}_\odot$ and half-mass radius $r_{\rm hm} = 1{\rm pc}$. The black dotted line represents the Plummer mass density profile of the host globular cluster.}
                \label{fig:rho_r_varN}
        \end{center} 
\end{figure*}

Supposing that the system is characterised by a length scale, $r_c$, we may introduce the dimensionless variables:
\begin{align}
        x &= \frac{r}{r_c},\;
        \tilde{M}_{\rm b}(x) = \frac{\mathcal{M}_{\rm b} (x)}{M_{\rm b}},\;
        \tilde{M}_\star(x) = \frac{\mathcal{M}_\star (x)}{M_\star},\\
                \tilde{p}_{\rm b}(x) &= 4\pi r_c^3 p_{\rm b},\;
                y(x) = -\ln \frac{\tilde{p}_{\rm b}(x)}{\tilde{p}_{\rm b}(0)}.
\end{align} 
The system (\ref{eq:ode_system_d1})-(\ref{eq:ode_system_d2}) becomes
\begin{align}
\label{eq:ode_system_1}
\frac{dy (x)}{dx} &= \frac{G m_{\rm b} M_\star}{r_c T_\star(x)}\frac{1}{x^2}  \left(\tilde{M}_\star(x)+\frac{M_b}{M_\star}\tilde{M}_{\rm b}(x)\right) + \frac{T_\star(x)^\prime}{T_\star(x)} 
,\\
\label{eq:ode_system_2}
\frac{d\tilde{M}_{\rm b}(x)}{dx} &=
 \tilde{p}_{\rm b}(0) x^2 e^{-y(x)},
\end{align}
with initial conditions
\begin{equation}\label{eq:ini_cond}
        y(0) = 0,\quad \tilde{M}_{\rm b}(0) = 0
        ,\quad
        y(0)^\prime = 0,\quad \tilde{M}_{\rm b}(0)^\prime = 0
\end{equation}
and subject to the boundary constraint
\begin{equation}\label{eq:bound_cond}
\tilde{M}_{\rm b}(\infty) = 1 \Leftrightarrow
\tilde{p}_{\rm b}(0) = \left(\int_0^\infty dx\, x^2 e^{-y(x)} \right)^{-1}.
\end{equation}
For given host distributions $\tilde{M}_\star(x)$, $T(x)$ the system (\ref{eq:ode_system_1})-(\ref{eq:bound_cond}) may be solved numerically.

We consider again the case where the stellar distribution is that of a Plummer sphere, as in Equation \ref{eq:Plummer}.
The temperature is identified as
$T_\star (r) = m_\star \sigma_\star(r)^2$.
The system (\ref{eq:ode_system_1})-(\ref{eq:ode_system_2}) becomes
\begin{align}
\label{eq:ode_system_Pl_1}
\frac{dy (x)}{dx} &= 6 \frac{m_{\rm b}}{m_\star}  \left( \frac{x}{1+x^2} + \frac{M_{\rm b}}{M_\star} \frac{\sqrt{1+x^2}}{x^2} \tilde{M}_b(x) \right) - \frac{x}{1+x^2} ,
\\
\label{eq:ode_system_Pl_2}
\frac{d\tilde{M}_{\rm b}(x)}{dx} &=
\tilde{p}_{\rm b}(0) x^2 e^{-y(x)},
\end{align}
subject again to the conditions (\ref{eq:ini_cond})-(\ref{eq:bound_cond}).

The reach (or otherwise) of an equilibrium and the specific form of the equilibrium distribution of the Brownian particles, which we consider hereof to be a BH population, depend on both the number of bodies and their individual mass. 
In Figure{\ref{fig:M_ratio_p0}, we  calculate the series of equilibria of BH populations immersed in a Plummer stellar profile for various ratios $m_{\rm b}/m_\star$. Different points of each curve correspond to the equilibrium state of a different BH population with total mass, $M_{\rm b}$, and corresponding central density $\tilde{p}(0)$. We identify an instability that sets in  at the maximum of the series of equilibria curve. No equilibria exist above the maximum, whereas, according to the Poincar\'e theorem of linear series of equilibria, the branch beyond the turning point (dotted lines in Figure \ref{fig:M_ratio_p0}) are unstable. 
        
        We further discover numerically that when the BH subcluster mass is scaled to become the Spitzer parameter,
        \begin{equation}
                S \equiv \frac{M_{\rm b} m_{\rm b}^{3/2} }{M_\star m_\star^{3/2}} 
        \end{equation}
and the central probability density is scaled as \begin{equation} 
B \equiv \tilde{p}_{\rm b}(0)\left(\frac{m_\star}{m_{\rm b}}\right)^{3/2},
\end{equation} 
then all curves with different ratios, $m_{\rm b}/m_\star,$ merge to form a single one, as in Figure \ref{fig:Spitzer} (we estimate the exact value of the exponent to be $1.51$ and not $3/2$, but we consider this small deviation a numerical precision effect without any physical significance).     Thus, the curve \ref{fig:Spitzer} is global and applies to all Brownian populations immersed in a Plummer profile. The instability sets in at
 \begin{equation}
         B_{\rm I} = 140, \;
         S_{\rm I} = 0.25. 
 \end{equation}
Any stationary equilibrium with
\begin{equation}
B > B_{\rm I}
\end{equation}
is unstable within the current framework.
In addition no stationary equilinbrium states exist above 
\begin{equation}
S > S_{\rm I}.
\end{equation}
This value $S_{\rm I}$ is significantly larger than the Spitzer value \citep{1969ApJ...158L.139S}, $S_{\rm SP} = 0.16,$ calculated for isothermal equipartition. 

We stress at this point that while we assume here that the BH population to be immersed inside a fixed density profile of the cluster, it should, in practice, affect the cluster profile at least in the vicinity of BH subcluster's denser regions. This effect, not taken into account in the current work, may influence the cluster's ability to support the BH population. Nevertheless, it seems possible that the BH population will locally heat up and inflate the population of lighter bodies in an effect that is similar to osmosis. This could possibly enhance, rather than reduce, a cluster's ability to support the BH population via gravitational fluctuations and could lead to formation of a core (as e.g., proposed by \citealt{2004ApJ...608L..25M}). Such BH population feedback onto the cluster density profile requires further investigation and is not studied here. 

The density profile of a BH population at the onset of the instability appears in Figure \ref{fig:rho_r_varN}, where the Plummer profile, $\rho_\star$, is also plotted . It is evident that the BH subcluster may extend up to $0.2{\rm pc}$ inside the host cluster and it may be almost twice as dense in the centre.

        \begin{figure}[!tb]
        \begin{center}
                \subfigure[Series of equilibria of BH-subcluster.]{
                        \label{fig:Nb_GC}
                        \includegraphics[scale = 0.5]{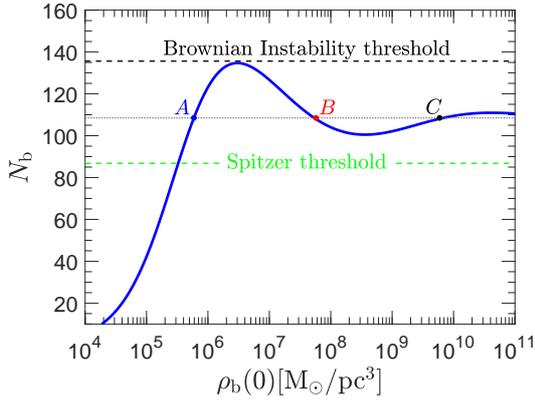}  }
                \subfigure[Density profile of equilibria $A$, $B$, $C$.]{
                        \label{fig:rho_r_GC}
                        \includegraphics[scale = 0.5]{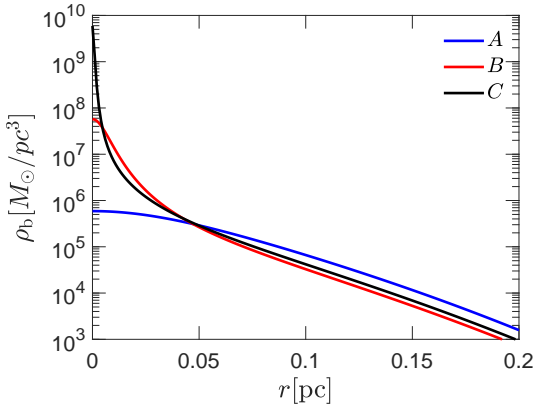}  }
                \caption{(a) Series of equilibria of BH populations with  individual BH mass, $m_{\rm b} = 10{\rm M}_\odot,$ expressed by the number of BHs, $N_{\rm b},$ with respect to the BH population central density, $\rho_{\rm b}(0)$. The BH population is immersed in a Plummer profile of a globular cluster with average individual stellar mass, $m_\star = 0.5{\rm M}_\odot$, number of stars, $N_\star = 10^6,$ and half-mass radius, $r_{\rm hm} = 1{\rm pc}$. For a BH subcluster with $S > 0.185,$ corresponding here to $N_{\rm b} > 100$,  many equilibrium solutions exist. We consider $S=0.2,$ that is, $N_{\rm b} = 109$ and the first three corresponding equilibria $A$, $B$, $C$. (b) Density profile of the three stationary states $A$, $B$, $C$ specified in (a). Profile $A$ is stable. The equilibria $B$ and $C$ are unstable with respect to Brownian fluctuations, although they may get stabilized by other processes. }
                \label{fig:Nb-rho_GC}
        \end{center} 
\end{figure}

\begin{figure}[!tb]
        \begin{center}
                \subfigure[$N_\star = 10^6$.]{
                        \label{fig:Nb_max_GC}
                        \includegraphics[scale = 0.5]{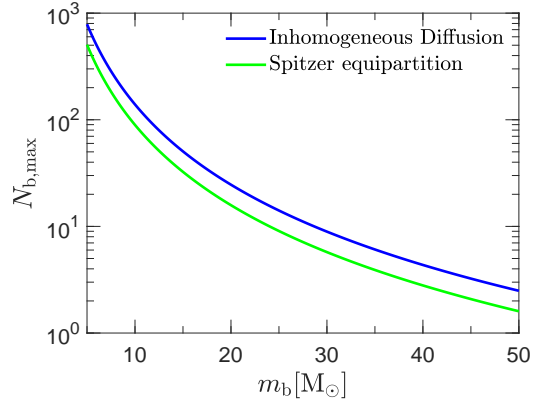}  }
                \subfigure[$N_\star = 10^8$.]{
                        \label{fig:Nb_max_NSC}
                        \includegraphics[scale = 0.5]{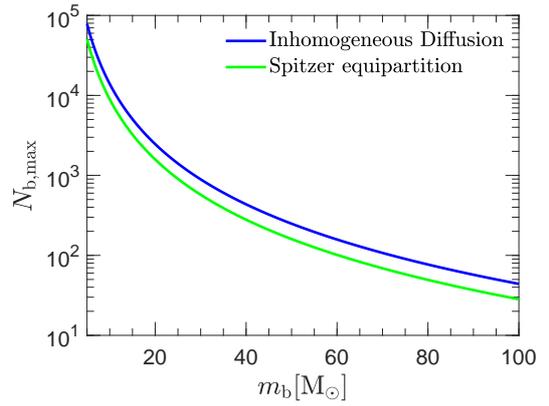}  }
                \caption{Maximum number of BHs with respect to their individual mass that may maintained at a stationary state inside a Plummer stellar profile with $m_\star = 0.5{\rm M}_\odot$ and two cases of a number of stars of $N_\star = 10^6  ,\, 10^8$. Upper blue curve corresponds to our model for inhomogeneous diffusion, while the lower green one to the Spitzer instability limit derived from the energy equipartition.}
                \label{fig:Nb_max}
        \end{center} 
\end{figure}

In Figure \ref{fig:Nb_max}, we plot the maximum number of Brownian bodies, that is, BHs, in our context and with respect to their individual average mass for inhomogeneous diffusion and Spitzer instability. We assume $m_\star = 0.5{\rm M}_\odot$ and consider two cases of $N_\star = 10^6,\, 10^8$. In the typical case of a globular cluster with $N_\star = 10^6$ and $m_{\rm b} = 10 {\rm M}_\odot,$  for an inhomogeneous diffusion we get $N_{\rm b} \sim 1400,$ while the Spitzer threshold is significantly lower at $N_{\rm b} \sim 890$.

        \begin{figure}[!tb]
        \begin{center}
                \subfigure[IMBH and equal mass BH subclusters.]{
        \label{fig:p_IMBH_subclusters}
        \includegraphics[scale = 0.5]{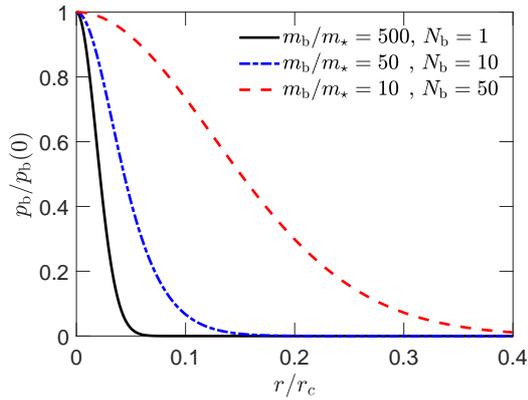}  }
                \subfigure[IMBH and different subcluster profiles.]{
                        \label{fig:p_IMBH_phases}
                        \includegraphics[scale = 0.5]{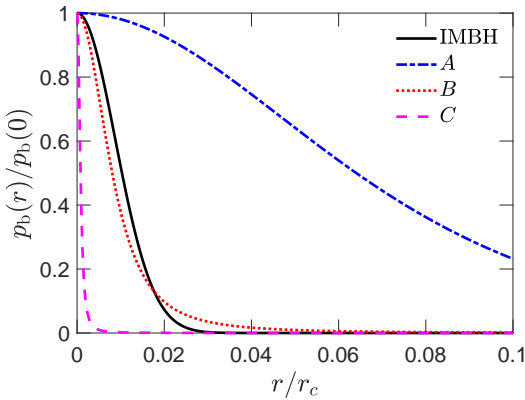}  }
                \caption{(a) Probability density distribution of a single massive BH (continuous curve) along with that of BH populations with equal total mass. It is assumed $N_\star = 10^6$. (b) A massive BH with $m_{\rm b}=1090{\rm M}_\odot$ (continuous line), along with the three different possible equilibria of Figure \ref{fig:Nb-rho_GC} corresponding to the same BH population with $N_{\rm b} = 109$, $m_{\rm b} = 10 {\rm M}_\odot$.  It is assumed $N_\star = 10^6$, $m_\star = 0.5{\rm M}_\odot$.  }
                \label{fig:IMBH-subclusters}
        \end{center} 
\end{figure}

In Figure \ref{fig:IMBH-subclusters}, we compare the probability distribution of a single intermediate-mass BH with BH populations of lighter stellar mass BHs but with equal total mass. It is evident from Figure \ref{fig:p_IMBH_subclusters} that in order to be able to discriminate a BH population from an intermediate-mass BH with a mass of $500 m_\star$, it may be required to probe inside the inner $0.1{\rm pc}$. In Figure \ref{fig:p_IMBH_phases}, we show that for a BH population with $N_{\rm b} = 1090$, $m_{\rm b} = 10{\rm M}_\odot$ the first unstable profile is very similar to that of an intermediate-mass BH of equal total mass, while in order to discover observationally the second unstable profile, if stabilized by other processes, one has to probe the inner $\sim 0.005{\rm pc}$.

\section{Conclusions}

In this work, we study inhomogeneous diffusion, that is, diffusion in a medium with varying mass density and temperature and, hence, also varying damping and diffusion coefficients. This is the typical case for self-gravitating systems that are spatially inhomogeneous and trapped in states with varying velocity dispersions. We argue that the inhomogeneous diffusion equation (\ref{eq:Brownian_diff_3D}) applies to self-gravitating systems that involve a sub-population of fewer and heavier bodies and describes gravitational Brownian motion. The corresponding stationary states are given in Equations \ref{eq:p_stationary} and \ref{eq:f_stat}.
 We calculated the spatial probability distribution function of a single Brownian particle immersed in a Plummer profile, as depicted in Figure \ref{fig:IMBH}. We estimate that a single intermediate-mass BH may wander as far as $\sim 0.05{\rm pc}$ in a typical Globular cluster, while a single typical stellar BH may wander even ten times farther. We validate the reality of inhomogeneous diffusion of BHs immersed in stellar clusters, expressed by Equation \ref{eq:Brownian_diff}, with the use of $N$-body simulations, as in Figure \ref{fig:simulation_dPdV}. The theoretical prediction is well-matched by the simulation results.

Applying our framework to a Brownian population of massive bodies (focusing on BHs) inside a stellar cluster, which follows a Plummer density profile, we identify an instability that sets in for Brownian populations with Spitzer parameter,  $S_{\rm I} = 0.25$, and a new global parameter, $B_{\rm I} = 140$, which depends on the central density of the Brownian population. This is depicted in Figure \ref{fig:Spitzer_M}. For $B>B_{\rm I}$, any stationary equilibrium state is unstable despite the fluctuations of the gravitational field. For $S>S_{\rm I},$ no stationary states exist. This is a manifestation of the Spitzer instability, reinterpreted as the inability of the cluster to support the sub-population of heavier bodies by gravitational fluctuations. The dependence of the onset of the instability on the individual mass ratios in such a framework arises naturally. Furthermore, since the ordinary Spitzer instability occurs at a lower value, $S_{\rm SP} = 0.16$, the inhomogeneous diffusion allows more massive BH populations to reach stationary states than one would expect from isothermal equipartition.

An important limitation of our model for BH populations in Section \ref{sec:BH_sub} regarding its physical applicability is that it does not take into account the feedback of the BH population to the gravitational potential of the host cluster. Such a limitation for analyses on the Brownian motion of a single massive BH was also noted in \citep{2007AJ....133..553M}. The situation is very similar since our BH population has the same mass of our assumed single massive BH and it turns out to have about the same spatial probability distribution. Still, it was our intention to neglect this feedback in order to quantify the effect solely of the gravitational fluctuations to the BH population and inspect their stabilizing efficiency. If anything, it seems plausible that 
 the BH population will locally heat up and inflate the population of lighter bodies, as was also suggested by  \cite{2004ApJ...608L..25M}. This will enhance, and not reduce, the cluster's ability to support the BH population via gravitational fluctuations and lead to the formation of a core. We remark that in his seminal work on Brownian motion, \cite{1905AnP...322..549E,einstein2011investigations} interpreted it precisely as a response to osmotic pressure. Osmosis involves the diffusion of the solvent particles (in our case the stars) in the region of the solute (in our case, the BH population), with result the inflation of the region containing the solute. 
Such BH population feedback onto the cluster density profile, including the possible reality of such a phenomenon as "gravitational osmosis," requires further investigation.
 Another future improvement could include following the time evolution of the diffusion equation (\ref{eq:Brownian_diff_3D}) for the BH population. 

In conclusion, the fact that a BH subcluster can be supported by random fluctuations of the gravitational field beyond the limit of Spitzer instability threshold supports the idea that globular clusters can retain a significant BH population.

\bibliography{2020_ROUPAS_GBM_BHsub}
\bibliographystyle{aa}

\appendix  

\section{Diffusion equation for gravitational Brownian motion}\label{app:diffusion}

We reproduce here the diffusion equation of Brownian motion inside an inhomogeneous medium with varying temperature, as was first proposed by  \cite{1988JPCS...49..673V}. We further argue that this diffusion equation describes gravitational Brownian motion. It applies to self-gravitating systems, which are not only inhomogeneous, but also are typically non-isothermal, in the sense that the velocity dispersion varies with position.

As suggested by  \cite{1943RvMP...15....1C}, the gravitational field, $\bm{g}(\bm{r},t),$ at a point, $\bm{r},$ and time, $t,$ of an $N$-body self-gravitating system may be decomposed to the sum of a mean field $\bm{g}_{\rm m}$ and a fluctuating field $\bm{g}_{\rm f:}$
\begin{equation}
\bm{g}(\bm{r},t) = \bm{g}_{\rm m}(\bm{r},t) + \bm{g}_{\rm f}(\bm{r},t).
\end{equation}
The mean field represents the effect of the system as a whole to each point of space at any instant of time through the smoothed out continuous mass density function $\rho(\bm{r},t):$
\begin{equation}
\bm{g}_{\rm m} = -\nabla \Phi,\quad 
\Phi(\bm{r},t) = -\int d\tilde{\bm{r}}\,
G\frac{\rho(\bm{r},t)}{|\bm{r}-\tilde{\bm{r}}|}.
\end{equation}
The fluctuating field accounts for the deviations from this smoothed out field that arise due to the granularity of the system. \cite{1943ApJ....97..255C} showed further that each body is subject to a dynamical friction force with a damping coefficient, $\eta,$ which is reciprocal to the relaxation timescale of the system.  Central to his derivation is a Fokker-Planck type of equation discovered by Kramers, namely,
\begin{equation}\label{eq:Kramers}
\left(  \frac{\partial }{\partial t}
+ v\frac{\partial }{\partial x} - \Phi^\prime\frac{\partial }{\partial v}
\right) f(x,v,t)
= \eta  \frac{\partial }{\partial v}
\left(v  +  \frac{T}{m}\frac{\partial }{\partial v}\right)f(x,v,t),
\end{equation}
which we consider here in one-dimension for simplicity without loss of generality. We denote $f$ as the phase-space probability distribution function, $m$  the mass of the Brownian particle, $T$ the temperature in units $k_{\rm B} = 1,$ and $\eta$ the damping coefficient.
Equation \ref{eq:Kramers} was derived by \cite{1940Phy.....7..284K} according to very general considerations. \cite{1943RvMP...15....1C} rederived alternatively Kramers equation and used it in self-gravitating systems to establish the concept of dynamical friction \citep{1943ApJ....97..255C}. Here, following \cite{1988JPCS...49..673V}, we consider Kramers equation, but with varying damping (dynamical friction in our case) coefficient and temperature,
\begin{equation}
\eta = \eta (x),\quad T = T(x).
\end{equation}
From Equation \ref{eq:Kramers}, we derive the diffusion equation, which is to be a generalization of the Smolukowski equation and we argue that it describes gravitational Brownian motion.

First, for the purposes of completeness, we reproduce the derivation of (\ref{eq:Kramers}) following \cite{1943RvMP...15....1C}. We assume that the motion of a Brownian particle is described by a Langevin equation,
\begin{equation}\label{eq:Langevin}
\frac{d v}{dt} = -\eta v + g_{\rm m} + g_{\rm f},
\end{equation}
where $\eta$ is the dynamical friction coefficient \citep{1943ApJ....97..255C}, $g_{\rm m} = -\Phi^\prime$ is the mean field and $g_{\rm f}$ the fluctuating field. 
The latter term has statistically defined properties rendering the Langevin equation a stochastic differential equation that cannot be solved as an ordinary differential equation might be. Instead, we would aim to derive a distribution function $f(x,v,t+\Delta t)$ governing the probability of occurrence of $x$,$v$ at time $t+\Delta t$ given the distribution, $f(x,v,t),$ at time, $t$.

We let $\Delta t$ be short with respect to the relaxation time $\Delta t \ll \eta^{-1}$ but long compared to the period of $g_{\rm f}$ fluctuations.
We define the velocity variation due to field fluctuation within $\Delta t $,
\begin{equation}\label{eq:B}
        {\rm B} (\Delta t) = \int_t^{t+\Delta t} g_{\rm f}(\xi) d\xi.
\end{equation}
Physically, it describes the net acceleration exerted on the Brownian particle arising from field fluctuations in time, $\Delta t$.
We impose the physical requirement that for $t\rightarrow \infty$, i.e. $t \gg \eta^{-1}$ that is at zeroth order of $\eta^{-1}$, the distribution function, $f,$ is locally Maxwellian. \cite{1943RvMP...15....1C} has proven, as in his Chapter II.2, that this requirement implies that ${\rm B}$
satisfies the Brownian probability distribution: 
\begin{equation}\label{eq:P_Brownian}
\psi({\rm B}) = \left( \frac{m}{4\pi \eta T \Delta  t}\right)^{1/2} e^{-\frac{m}{ 4 \eta T \Delta t}{\rm B}^2}.
\end{equation}

The spatial and velocity displacements, $\Delta x$, $\Delta v,$  at $\Delta t$ are derived by the Langevin equation (\ref{eq:Langevin}), so that we get 
\begin{equation}\label{eq:displacements}
\Delta x = v \Delta t,
\quad
{\rm B} = \Delta v + (\eta v + \Phi^\prime)\Delta t .
\end{equation}
The latter equation follows by integration of the Langevin equation (\ref{eq:Langevin}) from $t$ to $t+\Delta t$, introducing ${\rm B}$ by use of (\ref{eq:B}), and finally solving with respect to ${\rm B}$.

Given that Brownian motion is governed by two-body relaxation and it proceeds in a short timescale, it is reasonably expected to be modelled as a Markoff process\footnote{We caution the reader that the Markoff assumption may not be valid for any gravitational system. In the case of regular orbits, e.g. in clusters whose self-gravity is dominated by a central potential, the relaxation of orbital angular momentum may proceed in a different timescale than the energy and encounters between stars may be coherent, as in resonant relaxation \citep{1996NewA....1..149R}. There is recent evidence of resonant relaxation occuring in globular clusters, questioning also the local-scattering relaxation theory \citep{2019MNRAS.490..478L}.
as is pointed out also by \cite{1943ApJ....97..255C}.}  
In this case the probability distribution at any time, $t+\Delta t,$ can be derived from the distribution at previous time, $t$. 
Using Equation~\ref{eq:displacements}, the transition probability, $\psi(v-\Delta v;\Delta v),$ that $v$ changes by $\Delta v$ may be expressed with respect only to $\Delta v$. Then we have:
\begin{equation}\label{eq:Markoff}
f(x+v\Delta t,v,t+\Delta t) = \int d(\Delta v)\, f(x,v-\Delta v,t) \psi (x,v- \Delta v;\Delta v),
\end{equation}
where $\psi$ is given from (\ref{eq:P_Brownian}) for ${\rm B}(\Delta v)$ given in (\ref{eq:displacements}).
Taylor expanding $f(x+v\Delta t,v,t+\Delta t) $, $f(x,v-\Delta v,t)$ and $\psi (x,v- \Delta v;\Delta v)$ we get the Fokker-Planck equation:
\begin{equation}\label{eq:FP}
\left( \frac{\partial f}{\partial t} + v\frac{\partial f}{\partial x}\right) \Delta t + \mathcal{O}((\Delta t)^2) 
= -\frac{\partial (f \LA \Delta v\RA )}{\partial v} + \frac{1}{2}
\frac{ \partial (f\LA \Delta v^2\RA)}{\partial v^2} 
+ \mathcal{O}((\Delta t)^2) 
,\end{equation}
where the mean values $\LA \bullet\RA$ are calculated with the distribution $\psi(x,v;\Delta v)$. We get $\LA \Delta v\RA = -(\eta v + \Phi^\prime) \Delta t$ and $\LA (\Delta v)^2 \RA = (2\eta T/m)\Delta t + \mathcal{O}(\Delta t^2)$. Making a substitution in (\ref{eq:FP}), we get finally Kramers equation (\ref{eq:Kramers}), generalized for an inhomogeneous medium:
\begin{equation}\label{eq:Kramers_inh}
\left(  \frac{\partial }{\partial t}
+ v\frac{\partial }{\partial x} - \Phi(x)^\prime\frac{\partial }{\partial v}
\right) f(x,v,t)
= \eta (x)\frac{\partial }{\partial v}
\left(   v  + \frac{T(x)}{m}\frac{\partial }{\partial v}\right)f(x,v,t).
\end{equation}

Now, we derive the corresponding diffusion equation from Kramers equation following  \cite{1988JPCS...49..673V}, who uses a well-established method that was also used to derive the Smolukowski equation from the Kramers equation  \citep{1943RvMP...15....1C}, but modified in a straightforward manner to account for the dependence of $\eta$ and $T$ on $x$.
That is a method of elimination of fast variables \citep{1985PhR...124...69V} in which we expand the solution to (\ref{eq:Kramers_inh})     in powers of 
\begin{equation}
\varepsilon(x) \equiv \eta(x)^{-1},
\end{equation}
which is the local relaxation time \citep{1943RvMP...15....1C}.
We assume
\begin{equation}
f = f^{(0)} + f^{(1)} + f^{(2)} + \mathcal{O}(\varepsilon^3 ) 
\end{equation}
and get up to the second order the equations
\begin{align}
\label{eq:Kramers_exp_1}
&\frac{\partial }{\partial v}
\left( v  f^{(0)} + \frac{T}{m}\frac{\partial f^{(0)}}{\partial v}\right) = 0 \\
\label{eq:Kramers_exp_2}
&\varepsilon \left(\frac{\partial f^{(0)}}{\partial t}
+ v\frac{\partial f^{(0)}}{\partial x} - \Phi^\prime\frac{\partial f^{(0)}}{\partial v}
\right) = \frac{\partial }{\partial v}
\left( v  f^{(1)} + \frac{T}{m}\frac{\partial f^{(1)}}{\partial v}\right) \\
\label{eq:Kramers_exp_3}
& \varepsilon \left(\frac{\partial f^{(1)}}{\partial t}
+ v\frac{\partial f^{(1)}}{\partial x} - \Phi^\prime\frac{\partial f^{(1)}}{\partial v}
\right) = \frac{\partial }{\partial v}
\left( v  f^{(2)} + \frac{T}{m}\frac{\partial f^{(2)}}{\partial v}\right).
\end{align}
Requiring that $f^{(0)}$ vanishes for $|v|\rightarrow \infty$, the first equation gives
\begin{equation}\label{eq:f0_ansatz}
f^{(0)} = s(x,t) e^{-\frac{m v^2}{2T(x)}},
\end{equation}
for some function, $s(x,t),$ to be determined.
The integral of the right-hand side of Eq.~(\ref{eq:Kramers_exp_2}) is zero:
\begin{equation}
\int_{-\infty}^{+\infty} dv  \left( v  f^{(1)} + \frac{T}{m}\frac{\partial f^{(1)}}{\partial v}\right) =0
\end{equation}
and therefore the integral on the left-hand side should also be zero:
\begin{equation}\label{eq:lfs_int}
\int_{-\infty}^{+\infty} dv\,  
\frac{\partial f^{(0)}}{\partial t}
+ \int_{-\infty}^{+\infty} dv\,   v\frac{\partial f^{(0)}}{\partial x} 
- \int_{-\infty}^{+\infty} dv\, \Phi^\prime\frac{\partial f^{(0)}}{\partial v}
= 0
\end{equation}
Substituting Equation~\ref{eq:f0_ansatz} we have
\begin{align*}
&       \int_{-\infty}^{+\infty} dv\,   v\frac{\partial f^{(0)}}{\partial x} = s^\prime \int_{-\infty}^{+\infty} dv\,  v e^{-\frac{mv^2}{2T}}  + s\frac{ mT^\prime}{2 T^2} \int_{-\infty}^{+\infty} dv\,  v^3 e^{-\frac{mv^2}{2T}} = 0 
\\
& \int_{-\infty}^{+\infty} dv\, \Phi^\prime\frac{\partial f^{(0)}}{\partial v} = s  \Phi^\prime \int_{-\infty}^{+\infty} dv\, \frac{\partial}{dv}\left( e^{-\frac{mv^2}{2T}}\right) = 0 
\end{align*}
and therefore Equation~\ref{eq:lfs_int} gives the integrability condition:
\begin{equation}
\frac{\partial f^{(0)}}{\partial t} = 0 
,\end{equation}
which results finally in
\begin{equation}\label{eq:f0}
f^{(0)} = f^{(0)}(x,v) = s(x) e^{-\frac{m v^2}{2T(x)}}.
\end{equation}
We substitute this in Equation~\ref{eq:Kramers_exp_2} and get
\begin{align}
\varepsilon e^{-\frac{m v^2}{2T}}&\left( v s^\prime+ \frac{v}{T} m\Phi^\prime s + \frac{v^3}{2T^2} mT^\prime s \right) =\nonumber \\
&=
\frac{\partial }{\partial v}
\left( v  f^{(1)} + \frac{T}{m}\frac{\partial f^{(1)}}{\partial v}\right) 
= \frac{T}{m} \frac{\partial}{\partial v}\left(e^{-\frac{mv^2}{2T}} \frac{\partial}{\partial v} e^{\frac{mv^2}{2T}} f^{(1)}  \right).
\end{align}
The ansatz
\begin{equation}\label{eq:f1}
f^{(1)}(x,v,t) = \left( v h(x,t) + v^3 q(x,t) \right) e^{-\frac{mv^2}{2T}} 
\end{equation}
gives
\begin{align}
&h = h(x) =  -\frac{(T(x)s(x))^\prime + m\Phi(x)^\prime s(x)}{ T(x)}\varepsilon(x)
\\
&q = q(x) = -\frac{mT(x)^\prime}{6 T(x)^2}\varepsilon(x) s(x). 
\end{align}
Both $h$ and $q$ do not depend on time.
The general solution may be obtained by adding a solution of the homogeneous problem
\begin{equation}
f^{(1)}(x,v,t) = \varepsilon \left(-v \frac{T^\prime}{T} s -v  s^\prime -v \frac{ m\Phi^\prime }{T} s -v^3  \frac{ mT^\prime }{6T^2} s + w(t)   \right) e^{-\frac{m v^2}{2T}}, 
\end{equation}
where $\varepsilon$, $T$, $\Phi$,  $s$  depend only on $x$. We have 
\begin{equation}\label{eq:n_exp}
p(x,t) \equiv \int_{-\infty}^{+\infty} dv\, f(x,v,t) = \sqrt{2\pi\frac{ T(x)}{m}} \left(s(x) + \varepsilon(x) w(t) \right) + \mathcal{O}(\varepsilon^2),
\end{equation}
where $p$ is the probability density in space. For $N$ Brownian particles, $n=Np$ can be identified as their average number density.

Equation (\ref{eq:Kramers_exp_3}) gives the integrability condition
\begin{equation}
\int_{-\infty}^{+\infty} dv\, \frac{\partial f^{(1)}}{\partial t} = 
- \frac{\partial }{\partial x} \int_{-\infty}^{+\infty} dv\, v f^{(1)}.  
\end{equation}
Since $\partial f^{(0)}/\partial t = 0,$ we have
\begin{align}
\frac{\partial p(x,t)}{\partial t} &=
- \frac{\partial }{\partial x} \int_{-\infty}^{+\infty} dv\, v f^{(1)} + \mathcal{O}(\varepsilon^2)
\nonumber \\
&=   \frac{\partial }{\partial x} \left\lbrace \varepsilon  \left( \frac{\partial}{\partial x}(\frac{T}{m} \sqrt{2\pi \frac{T}{m}}(s + \varepsilon w)) + \Phi^\prime \sqrt{2\pi \frac{T}{m}}(s+\varepsilon w) \right) \right\rbrace  
\nonumber \\
&+ \mathcal{O}(\varepsilon^2),
\end{align}
which gives, by substitution of (\ref{eq:n_exp}), to first order in $\varepsilon,$ the diffusion equation
\begin{equation}\label{eq:Brownian_diff_app}
\frac{\partial p (x,t)}{\partial t} = \frac{\partial }{\partial x} \left\lbrace \mu (x)  \left( m\Phi(x)^\prime p(x,t)  + \frac{\partial}{\partial x}(T(x) p (x,t)) \right)\right\rbrace,
\end{equation}
where we denote $\mu(x) = \varepsilon(x)/m = 1/m\eta(x)$. This is the Van Kampen inhomogeneous diffusion equation that we suggest applies to gravitational Brownian motion.

\end{document}